\documentclass[12pt]{iopart}
\newcommand{\be}{\begin{equation}}
\newcommand{\ee}{\end{equation}}

\newcommand{\bea}{\begin{eqnarray}}
\newcommand{\eea}{\end{eqnarray}}
\newcommand{\half}{\frac{1}{2}}

\def\al{\alpha}

\def\om{\omega}

\def\lab{\label}

\begin{document}

\title[Time-reversal violation as loop-antiloop symmetry breaking and
the Bessel equation]{Time-reversal violation as loop-antiloop symmetry breaking:
the Bessel equation, group contraction and dissipation}

\author{Eleonora Alfinito\dag\ and Giuseppe Vitiello\ddag \footnote[3]{To whom
correspondence should be addressed (vitiello@sa.infn.it)}}

\address{\dag\ INFM, Sezione di Lecce, 73100 Lecce, Italy}

\address{\ddag\ Dipartimento di Fisica ``E.R.Caianiello",
Universit\`a di Salerno, 84080
Salerno, Italy\\
INFN, Gruppo Collegato di Salerno and INFM, Sezione di Salerno}

\eads{\mailto{eleonora.alfinito@unile.it},
\mailto{vitiello@sa.infn.it}}



\begin{abstract}

We show that the Bessel equation can be cast, by means of suitable
transformations, into a system of two damped/amplified parametric
oscillator equations. The relation with the group contraction
mechanism is analyzed and the breakdown of loop-antiloop symmetry
due to group contraction manifests itself as violation of
time-reversal symmetry. A preliminary discussion of the relation
between some infinite dimensional loop-algebras, such as the
Virasoro-like algebra, and the Euclidean algebras $e(2)$ and
$e(3)$ is also presented.

\end{abstract}



\newpage

\section{Introduction}


In this paper we show that the Bessel equation can be cast, by
means of suitable transformations, into a system of two parametric
oscillator equations, one for a damped oscillator, the other one
for an amplified one. We show that the group contraction mechanism
\cite{inonu} is involved in such a relation of the Bessel equation
with the dissipation/amplification system.

The interest in such a representation of the Bessel equation is
due to the role played by the couple of damped/amplified
oscillators in several physical problems \cite{Dekker}. Since
Bateman  \cite{bateman} proposed to treat the damped system by
doubling its degrees of freedom and introducing the companion
(time-reversed) amplified system, the interest in the
damped/amplified oscillator system has been growing \cite{FT} and
its relevance in treating dissipation at classical and quantum
level has been stressed \cite{QD} - \cite{blasoneJ}. The couple of
damped/amplified oscillators has been fruitfully used to describe
inflationary models of the Universe  \cite{AV1}, thermal field
theories \cite{QD}, Chern-Simons gauge theory  \cite{BPV}, Bloch
electrons in metals  \cite{BPV}, the dissipative quantum model of
brain \cite{V,AV2,MyD}, etc. It also presents features common to
the formalism of two-dimension gravity models \cite{jackiw}. In
particular, the equivalence, under suitable parametrization,
between the spherical Bessel equation and the damped/amplified
oscillators was firstly recognized in the study of expanding
geometry models  \cite{AV1} and of ordered domain formation in
brain models  \cite{AV2}.

On the other hand, the possibility of using and of exploiting the
properties of special functions in physical problems is {\it in
itself} of great interest, since, as remarked by Wigner
 \cite{talman}, "the role which is common to all special functions
is to be matrix elements of representations of the simplest Lie
groups". There is therefore a strong motivation to analyze the
relation between the Bessel equation and the damped/amplified
parametric oscillator system also from the perspective of special
function theory and group theory. We show that the mechanism of
group contraction \cite{inonu} by which one gets the Euclideans
groups $E(2)$ and $E(3)$, whose representations are given in terms
of planar and spherical Bessel functions, respectively, introduces
the breakdown in the loop-antiloop symmetry around a preferred
axis. In turn, this can be read off, in a given
re-parametrization, as the breakdown or violation of the
time-reversal symmetry.

The analysis in this paper brings us also to consider infinite
dimensional loop-algebras, such as the Virasoro-like algebra, in
relation with the Euclidean algebras $e(2)$ and $e(3)$. In view of
the relevance of infinite dimensional algebraic structures in many
physical problems, the preliminary results here presented deserve
to be further analyzed, which is our plan for a future work.

The plan of the paper is the following. In Section 2 we consider
the Bessel equation for the spherical Bessel functions and derive
from it the set of two damped/amplified parametric oscillators. In
Section 3 we present similar derivation for the Bessel equation
for the planar Bessel functions and we discuss the role played by
the group contraction in our derivation. In Section 4 some
preliminary results on the relation between the Virasoro-like
algebra and the $e(2)$ and $e(3)$ algebras are presented. Section
5 is devoted to conclusions and further remarks. In the Appendix
we show that the Bessel-like equation describes the damped
parametric oscillator under suitable conditions.

\section{The Bessel equation and dissipation}

In this Section we show that the spherical Bessel equation of
order $n$ (also called the Bessel equation of fractional order
 \cite{Abramowitz}) can be cast, by means of suitable
transformations, into a set of two equations representing a couple
of damped/amplified parametric oscillators. In the following
Section we analyze the case of the planar Bessel equation and the
group structure underlying the relation between Bessel equation
and damped/amplified oscillators.

The spherical Bessel equation of order $n$, whose solutions
constitute a complete set of (parametric) decaying functions
 \cite{Abramowitz}, is:
\be \eta^2 \ J_{n;\eta \eta} + 2\eta \ J_{n;\eta} + [\eta^2 -
n(n+1)] \ J_{n} \ = \ 0 ~. \label{qm42}\ee
Here $n$ is an integer or zero number, $(n = 0, \pm 1, \pm 2,
...)$ and, as customary, the labels ``$;\eta$" and ``$;\eta \eta$"
denote first and second order derivatives, respectively. As well
known, the solutions of  Eq. (\ref{qm42}), the so called spherical
Bessel functions, can be expressed in terms of  the first and
second kind Bessel functions and their linear combinations (the
Hankel functions).

Eq. (\ref{qm42}) is invariant under the transformation $n
\rightarrow -(n+1)$ and $J_{n}$ and $J_{-(n+1)}$ are both
solutions of the same equation. We can express this by regarding
(\ref{qm42}) as an eigenvalue equation and saying that $J_{n}$ and
$J_{-(n+1)}$ are degenerate solutions corresponding to the same
eigenvalue $n(n+1)$ of the operator ${\eta}^2 \ {d^{2} \over {d
\eta}^{2}} + 2 \eta \ {d \over {d \eta}} + {\eta}^2$.

We now perform in the Eq.(\ref{qm42}) the change of variables :
$\eta \rightarrow \eta \equiv \epsilon x$ with $x \equiv
e^{-t/\al}$, where $\epsilon$ and $\al$ are arbitrary parameters
and the new variable $t$ may be thought to denote, e.g., the time
variable. By using $w_{n,l} \equiv J_{n}\ \cdot(x)^{-l}$,
Eq.(\ref{qm42}) then goes into the following one:
\be \fl
\stackrel{..}w_{n,l} \ - \ \frac{2l+1}{\al}\ \stackrel{.}
w_{n,l} + \left[\frac{l(l+1)-n(n+1)}{\al^{2}} \ +\   (
\frac{\epsilon}{\al})^{2}\ e^{-2t/\al} \right] \ w_{n,l}\ = \ 0,
\label{qm7}\ee
where $\stackrel{.} w$ denotes derivative of $w$ with respect to
time $t$. We now remark that making the choice $l(l+1)=n(n+1)$ the
degeneracy between the solutions $J_{n}$ and $J_{-(n+1)}$ is
removed: in other words, a {\it partition} is induced between the
two solution sectors $\{J_{n}\}$ and $\{J_{-(n+1)}\}$, as shown by
the fact that now two {\it different} sets of equations are
obtained, one for ${ w}_{n,l}$ and the other one for ${
w}_{-(n+1),l }$, respectively, each set being composed by two {\it
different} equations, one for $l=-(n+1)$ and the other one for
$l=n$. The set for ${ w}_{n,l}$ is
\bea &\stackrel{..}{ w}_{n, -(n+1)} \ + \ \frac{2n+1}{\al} \
\stackrel{.}{w}_{n, -(n+1)} + [( \frac{\epsilon}{\al})^{2}\
e^{-2t/\al} ] \ w_{n, -(n+1)}\ = \ 0,
\nonumber\\
&\stackrel{..}{ w}_{n,n} \ - \ \frac{2n+1}{\al}\ \stackrel{.}{
w}_{n,n} + [( \frac{\epsilon}{\al})^{2}\ e^{-2t/\al} ] w_{n,n}\ =
\ 0 ~, \label{qm6}\eea
for $l=-(n+1)$ and for $l=n$, respectively. Similarly, two
equations for ${ w}_{-(n+1),l}$ are obtained:
\bea \fl
&\stackrel{..}{w}_{-(n+1), n} - \frac{2n+1}{\al}
\stackrel{.}{w}_{-(n+1), n} + [(\frac{\epsilon}{\al})^{2}
e^{-2t/\al} ]w_{-(n+1), n}=0,
\nonumber\\
\fl
 &\stackrel{..}{w}_{-(n+1),-(n+1)} + \frac{2n+1}{\al}
\stackrel{.}{w}_{-(n+1),-(n+1)} + [(\frac{\epsilon}{\al})^{2}
e^{-2t/\al}]w_{-(n+1),-(n+1)} =  0 , \label{qm6a}\eea
for $l=n$ and $l=-(n+1)$, respectively. Inspection of Eqs.
(\ref{qm6}) and Eqs. (\ref{qm6a}) shows that the symmetry under
the transformation $n \rightarrow -(n+1)$ has been broken.

As a further step, let us now choose the arbitrary parameters
$\al$ and $\epsilon$ to be $n$-dependent: $\al \rightarrow
\al_{n}$ and $\epsilon \rightarrow \epsilon_{n}$ (such a choice
means that we use $\eta \rightarrow \eta_{n} \equiv \epsilon_{n}
x_{n}$ with $x_{n} \equiv e^{-t/\al_{n}}$ ). Also, we perform our
choice in such a way that $\frac{2n+1}{\al_{n}} \equiv L$ and
$\frac{\epsilon_{n}}{\al_{n}} \equiv \om_0$ do not depend on $n$
(and on time). By setting $u_{n} \equiv { w}_{n,-(n+1)}$, and
$v_{n} \equiv { w}_{n,n}$, we then see that Eqs. (\ref{qm6}) are
nothing else than the couple of equations for the damped/amplified
parametric oscillators:
\bea \stackrel{..}u_{n}\,+\,L \stackrel{.}u_{n}\,+\,
{\omega_{n}}^{2}(t) u_{n} \,= &\ 0,
\nonumber \\
\stackrel{..}v_{n}\,-\,L \stackrel{.}v_{n}\,+\,
{\omega_{n}}^{2}(t) v_{n} \,= &\ 0, \label{qm3}\eea
with frequency
\be
  \omega_{n}(t)\ =\ {\omega}_{0} \ e^{-\frac{L t}{2n+1}}.
\label{qm4} \ee
Eqs. (\ref{qm3}) are sometimes called Hill-type equations
 \cite{HT}. We see that $\omega_{n}(t)$ approaches to the
time-independent value $\om_{0}$ for $n \rightarrow \infty$: the
frequency time-dependence is thus "graded" by the order $n$ of the
original Bessel equation. $L$ and $\om_{0}$, which may be
arbitrarily chosen, are characteristic parameters of the
oscillator system.

We note that our choice of keeping $L$ independent of $n$ implies
that $\al_{-(n+1)}=-\al_{n}$. Then the transformation $n
\rightarrow -(n+1)$ leads to solutions (corresponding to
$J_{-(n+1)}$) which have frequencies exponentially increasing in
time (cf. Eq. (\ref{qm4})). These solutions can be respectively
obtained from the ones of Eqs. (\ref{qm3}) by time-reversal $t
\rightarrow -t$ and exchanging $u$ with $v$ (which we refer to as
"charge conjugation"). In the large $n$ limit ($\omega_{n}
\rightarrow \omega_{0}$) $u_{n}$ and $v_{n}$ are each the
time-reversed of the other one and in that limit the two sectors
$\{ J_{i} \}$, $i = n, -(n+1)$ are mapped one into the other one.

We thus finally recognize the core of the relation between the
spherical  Bessel equation and the dissipation/amplification
phenomenon: the breakdown of the $n \rightarrow -(n+1)$ symmetry
of the original spherical Bessel equation corresponds in our
representation to the breakdown of time-reversal symmetry (the
emergence of the {\it arrow of time}) in the manifold of the
solutions (the spherical Bessel functions) $\{ J_{i} \}$, $i = n,
-(n+1)$.

In the following Section we will see that such a feature has its
root in the structure of the $E(3)$ group, whose representations,
as well known, can be indeed constructed by means of the spherical
Bessel functions. We will also see that the breakdown of the $n
\rightarrow -(n+1)$ symmetry can be viewed as the breakdown of the
loop-antiloop symmetry around a preferred direction in the
4D-space.

We observe that the functions $w_{n,n}$  and $w_{n,- (n+1)}$ are
"harmonically conjugate" functions in the sense that they may be
re-conduced to the single parametric oscillator
\be \stackrel{..}{r}_{n} +\Omega_{n} ^2(t)r_{n}=  0 ~,
\label{qmm5} \ee
where $r_{n}(t)$ is defined through $w_{n,-(n+1)}(t) = {1\over
{\sqrt 2}} r_{n}(t)e^{{-L t\over 2}}$  and
 $w_{n,n}(t) = {1\over {\sqrt 2}} r_{n}(t)e^{{L t\over2}}$. $\Omega_n$
is the common frequency: \be
  \Omega_{n}(t) = \left [ \left ({\omega}_{n}^{2}(t)  - {{L
  ^{2}}\over{4}} \right ) \right ]^{1\over{2}}.
  \label{qm5} \ee

Remarkably, the first of Eqs. (\ref{qm3}), with $n = 1$ is
commonly used in expanding geometry (inflationary) models of the
Universe  \cite{AV1}. In that case $L$ denotes the Hubble
constant. For a discussion of expanding geometry models in terms
of Eqs. (\ref{qm3}) see  \cite{AV1}.

In conclusion, the spherical Bessel equation represents,  under
suitable transformations, a two-fold hierarchy of couples of
parametric oscillators with constant damping/amplification given
by $L$ and with time dependent frequency graded by $n$ and by
$-(n+1)$. Transition from one tower to the other one ($n
\rightarrow -(n+1)$) is induced by time-reversal $t \rightarrow
-t$ combined with "charge conjugation" $u \rightarrow v$.

In the Appendix we show that the Bessel-like equation
\be J^{''}  \ +   \frac{\alpha}{\eta} \ J^{'}  +   \left(1 -
\frac{\beta^{2}}{\eta^{2}} \right)\ J = \ 0 ~, \label{6.1a}\ee
where $\alpha=1$ or $2$, $\beta$ is an arbitrary real constant and
$^\prime$ indicates the derivative with respect to the independent
variable $\eta$, may represent parametric oscillators with
constant damping and with different functional choices for the
frequency. Vice-versa, from the equation of the damped parametric
oscillator, by use of convenient transformations, one can always
obtain the Bessel-like equation (\ref{6.1a}).

In the next Section we consider the group structure underlying the
planar and the spherical Bessel equation in order to clarify their
connection with the damped/amplified parametric oscillator system.

\section{Group contraction and Bessel equations}\label{s2}

It is well known that the representations of the Euclidean groups
can be constructed in terms of the Bessel functions
\cite{talman,Abramowitz}. We are therefore interested in these
groups in the present Section. Features analyzed in the previous
Section will be found to be rooted in these groups.

We will focus our attention on $E(2)$ and on $E(3)$, the Euclidean
groups in the plane and in the space, respectively. Their relation
with the Laplace equations in three and four dimensions will be
summarized (see \cite{talman} for details) in view of the
connection with Bessel equations.

\subsection{$E(2)$ and planar Bessel equations}\label{s2.1}

$E(2)$ is the group of the $T(\vec{v})R(\theta)$ transformations,
where $T(\vec{v})$ is the translation in the plane by the vector
$\vec{v}$  ($\vec{v} \equiv (a,b)$) and $R(\theta)$ is the
rotation of the plane around the origin by the angle $\theta$. The
associated Lie algebra is given in terms of the two translation
generators $P_{a},P_{b}$ and of the rotation generator M:
\be [P_{a},P_{b}]=0,\quad [P_{a},M]= - P_{b}, \quad [P_{b},M]=
P_{a}. \label{1.1}\ee
The invariant operator of  $E(2)$ is $P^{2}=P_{a}^{2}+P_{b}^{2}=$
$P_{+} \ P_{-}$=$P_{-}\ P_{+}$, with $P_{\pm} \equiv P_{a} \pm
iP_{b}$, which has non-positive eigenvalue, $-p^{2}$.

The group transformations on square integrable functions $f$ are
represented by the action of the operator $D(\vec{v},\theta)$ (see
e.g. \cite{talman}):
    \bea
    D(\vec{v},0)f(\phi) \,= &\,e^{i\vec{p}\cdot \vec{v}} f(\phi) , \nonumber \\
    D(0,\theta)f(\phi) \,= &\, f(\phi-\theta)~.
    \label{1.111}
    \eea
In particular, it is possible to assume as basis functions the
complete set of normalized eigenfunctions of the rotation subgroup
$\left\{f_{n}(\phi)\right\}, \quad f_{n} \equiv (2\pi)^{-\half}\
i^{-n}\ e^{in\phi}$: $D(0,\theta)f_{n}(\phi) \,= \,
e^{in\theta}f_{n}(\phi)$. The complete representation of $E(2)$
can be expressed in the form:
\bea
& D(\vec{v},\theta)f_{n}=\sum_{m}
\Delta(\vec{v},\theta)_{mn} f_{m},\nonumber\\
& \Delta(\vec{v},\theta)_{mn} = (-1)^{m-n} e^{-im\beta}\
J_{m-n}(pr) e^{in(\beta-\theta)}, \label{1.3}\eea
where $(r,\beta)$ are the polar coordinates of $\vec{v}$ and
$J_{m}$ is the planar Bessel function of order m. Eqs. (\ref{1.3})
provide the representation of $E(2)$ in terms of the Bessel
functions $J_{m}$ \cite{talman}.

Now we want to study the $P^2$ eigenvalue equation. To this aim we
consider the 3D-Laplace equation:
\be \nabla^{2}\psi \equiv \left(\frac{\partial^{2}}{\partial
x_{1}^{2}} + \frac{\partial^{2}}{\partial x_{2}^{2}}
+\frac{\partial^{2}}{\partial x_{3}^{2}}\right)\psi= 0,
\label{1.4}\ee
which refers to an isotropic and homogeneous 3D-space. By
choosing, instead of the spherical or rectangular coordinates, the
cylindrical ones, we have:
\be \left(\frac{\partial^{2}}{\partial r^{2}}
+\frac{\partial}{r\partial r} +\frac{\partial^{2}}{r^{2}\partial
\theta^{2}} +\frac{\partial^{2}}{\partial x_{3}^{2}}\right)\psi=
0, \label{1.5}\ee
and we search for solutions of the type:
\be
\psi(r,\theta,x_{3})=
\varphi(r,\theta)\cdot \sigma(x_{3}), \qquad
\frac{\partial^{2}}{\partial x_{3}^{2}}\sigma(x_{3})\equiv
p^{2}\sigma(x_{3}). \label{1.6}\ee
Square integrable eigenfunctions (for $p$ positive) are obtained
by
selecting, for positive $x_{3}$, the solution: $\sigma=
e^{-x_{3}p}$ and for negative  $x_{3}$ the solution: $\sigma=
e^{x_{3}p}$.

It is clear that the choice of the cylindrical, instead of the
spherical coordinates, breaks the symmetry of 3D-spatial rotation
group $SO(3)$: indeed, when cylindrical coordinates  are chosen
$x_{3}$ is differently treated  with respect to the two remaining
coordinates and this singles out a privileged axis for rotations.
The resulting symmetry group is $E(2)$, the group contraction of
$SO(3)$. This is manifest in the fact that the Laplace equation
(\ref{1.5}) reduces to the eigenvalue equation for $P^{2}$
(realized in polar coordinates), i.e. the 2D-Helmholtz equation:
\be P_{+}\ P_{-}\  \varphi = \left(\frac{\partial^{2}}{\partial
r^{2}} +\frac{\partial}{r\partial r}
+\frac{\partial^{2}}{r^{2}\partial \theta^{2}}\right)\varphi = -
p^{2} \varphi.
\label{1.7}\ee
If we solve this equation by assuming $\varphi(r,\theta) =
f(r)\cdot e^{in\theta}$, we obtain $f(r)=J_{n}(pr)$,
being $J_{n}(pr)$ the solution of the planar Bessel equation of
order $n$:
\be
 J_{n;\eta\eta} + \frac{1}{\eta} \ J_{n;\eta} + [1  -
 \frac{n^{2}}{\eta^{2}}]
\ J_{n} \ = \ 0, \label{1.8}\ee
where $\eta=pr$. Positive/negative values of $n$ correspond to
positive/negative rotations (loop/antiloop) around the $x_{3}$
axis, i.e., they correspond to different orientations of the
$x_{3}$ axis. The related solutions are, therefore, different
\cite{Abramowitz}. This occurrence (and the similar one in the
case of spherical Bessel functions studied below in subsection
3.2) bring us in a natural way to consider topological properties
of Bessel functions related with loop operators and loop-algebras,
on which we comment in Section 4. Notice that, in spite of the
fact that the solutions are not symmetric under the reversal of
the $x_{3}$ axis (i.e. under time-reversal in our choice $x_{3}
\equiv t$) Eq. (\ref{1.8}) is invariant under the $n \rightarrow
-n$ exchange. The $n \rightarrow -n$ invariance of Eq. (\ref{1.8})
actually reflects the existence of the two sets of the $SO(3)$
independent representations: the  $D^{n}$ and the $D^{-n-1}$
\cite{wyb}.

Now, by using a procedure similar to the one followed in Section 2
for the
spherical Bessel equation (\ref{qm42}), by setting $w_{n,l}(\eta)=
(\eta)^{-l} J_{n}(\eta)$, $\eta=\epsilon e^{-t/\al}$, and by
introducing the mirror parameter $l=\pm n$, the planar Bessel
equation (\ref{1.8}) gives the couple of damped/amplified harmonic
oscillators:
\bea l=-n:  & \stackrel{..}w_{n,-n} \ + \ \frac{2n}{\al}\
\stackrel{.} w_{n,-n} + \left[ ( \frac{\epsilon}{\al})^{2}\
e^{-2t/\al} \right] \ w_{n,-n}\
= \ 0,\nonumber \\
l=n:  & \stackrel{..}w_{n,n} \ - \ \frac{2n}{\al}\ \stackrel{.}
w_{n,n} + \left[ ( \frac{\epsilon}{\al})^{2}\ e^{-2t/\al} \right]
\ w_{n,n}\ = \ 0, \label{1.141}\eea
where the value of $l$ is connected with the choice of
negative/positive eigenvalues of the rotation generator $M$.

Summarizing, we have shown that the breakdown of the rotational
symmetry of $SO(3)$ which leads to its group contraction $E(2)$
introduces a crucial difference (loss of loop-antiloop symmetry,
indeed) in the double choice of the $x_{3}$ axis orientation.
This, in turn, results in the difference between the planar Bessel
functions or order $+ n$ and the ones of order $- n$, in terms of
which the $E(2)$ representations can be built. Then, we have shown
that the planar Bessel equation (\ref{1.8}) can be cast, by a
convenient reparametrization, into the set of eqs. (\ref{1.141})
for the damped/amplified harmonic oscillators: the {\it mirror}
index $\pm n$ of the Bessel functions is thus associated to the
{\it couple} of damped/amplified harmonic oscillators (it is a
time-mirror index).

It is also worth to recall that the contraction of $SO(3)$ to
$E(2)$ can be geometrically depicted as the projection of the
$S_{2}$ sphere (corresponding to $SO(3)$) on the plane (e.g. the
plane tangent to one of the poles of the sphere). It is then clear
that the radius, say $\rho$, of the sphere acts as a "scale" (one
may introduce $J_{x} \equiv \rho\cdot P_{a}, ~ J_{y} \equiv \rho
\cdot P_{b}, ~ J_{z} \equiv M$ with the $J$s denoting the
generators of the algebra $so(3)$): the $E(2)$ translations in the
tangent plane are "good" approximations in the limit  $\rho
\rightarrow \infty$, namely for distances much smaller than
$\rho$. The physical meaning of this is that the $SO(3)$
contraction to $E(2)$ manifests itself in local observations.
However, in the local observation process the $x_{3}$ axis
orientation is "locked". Which amounts to the loss of symmetry of
the solutions under the $n \rightarrow -n$ exchange (breakdown of
the loop-antiloop symmetry). As a matter of fact, specifying the
direction of the $x_{3}$ axis, i.e. choosing one of the two
possible forms for $\sigma$ (cf. Eq. (\ref{1.6})), produces
topologically inequivalent configurations \cite{mekhfi} (see also
Section 4).

We also observe that, identifying $x_{3}$ with the time $t$
coordinate, $p$ has the dimensions of an energy over an action
(cf., e.g., the eigenfunctions $\sigma= e^{\pm
x_{3}p}$).

As a final comment, we note that the so called ``spinor''
representation of $SO(3)$, which involves half-integer values of
$n$ (corresponding to represent $SO(3)$ in terms of the Pauli
matrices $\sigma_{i}, i=1,2,3$), maps the planar Bessel equation
into the harmonic oscillator equation. In fact, for $n=\pm\half$
and with $ w \equiv \sqrt{\eta}\ J_{\pm \half}$, Eq.(\ref{1.8})
reduces to the harmonic oscillator equation with frequency $p$ ($r$
plays the role of time):
\be \frac{d^{2} w(r)}{dr^{2}} + p^{2} \ w(r)=0,
  \qquad r=\frac{\eta}{p}.
\label{1.11a}\ee
In this connection it has been observed  \cite{Martin} that the
oscillator so obtained may be thought as a possible classical
analogue of a Fermi oscillator based on a 'rotation system'.

\subsection{$E(3)$ and the spherical Bessel equation} \label{s2.2}

The analysis presented in the previous subsection can be extended
to $E(3)$, the Euclidean group in the space, which is the group
contraction of SO(4). The algebra  $e(3)$ has six generators
$P_{i}$ and $M_{i}$, $i=1,2,3$, corresponding to the translation
and to the rotation generators, respectively. The commutation
relations are:
\be [P_{i},P_{j}]=0, \quad [M_{i},M_{j}]= \epsilon_{ijk}M_{k},
\quad [P_{i},M_{j}]=\epsilon_{ijk}P_{k}; \label{1.9}\ee
We observe that in the contraction process the $SO(3)$ subgroup
generated by the $M_{i}'s$  is left unchanged. The algebra $e(3)$
has two invariants, $P^{2} = \Sigma P_{i}^{2}$ and $\Sigma
P_{i}\cdot M_{i}$.

In the $4D$-space the Laplace equation for the function $\psi=$
$\psi(x_{1},x_{2},x_{3},x_{4})$ may be solved by using the
position: $\psi=\varphi(r,\theta,\phi)\cdot\sigma(x_{4})$,
 ($r,\theta,\phi$ spherical coordinates, $\sigma= e^{\pm
x_{4}p}$ ). By a procedure similar to
the one of the previous subsection, $x_{4}$ may be considered to
play the role of time $t$. The resulting equation (corresponding
to (\ref{1.7})) is solved by the function $\varphi=
Y_{n,m}(\theta,\phi)\cdot J_{n}(pr)$ where $Y_{n,m}$ is the
spherical harmonics and $J_{n}$ is the solution of the spherical
Bessel equation:
\be
 J_{n;\eta\eta} + \frac{2}{\eta} \ J_{n;\eta} + [1  -
 \frac{n(n+1)}{\eta^{2}}]
\ J_{n} \ = \ 0, \label{1.10}\ee
As in the previous $E(2)$ case for the planar Bessel functions,
the spherical Bessel functions depend on the continuous eigenvalue
$p^{2}$ of $P^{2}$ and are labelled by $n$ which is related with
the discrete eigenvalue $n(n+1)$ of the rotation operator ${\bf
M}^{2}$.

The order $n$ (integer or zero) classifies the representations
$D^{n}$ of the compact subgroup $SO(3)$ of $E(3)$. Actually, as
already said for the planar case, the existence of two sets of
$SO(3)$ independent representations (the  $D^{n}$ and the
$D^{-n-1}$)  \cite{wyb} is reflected in Eq.(\ref{1.10}) through
its invariance under the transformation $n \rightarrow -(n+1)$.
For $n=0$, or $n=-1$ equation (\ref{1.10}) reduces to the harmonic
oscillator equation with frequency $p$:
\be \frac{d^{2} w(r)}{dr^{2}} + p^{2} \ w(r)=0,
  \qquad r=\frac{\eta}{p},
\label{1.11}\ee
where $ w = \eta \ J_{0} \  {\rm or} \ \eta\ J_{-1}.$ The harmonic
oscillator thus appears to be related to the ``ground state'' (in
the $D^{n}$ or $D^{-n-1}$ spectra ) of the Bessel system. Here, of
course, the so-called ``true'' representation of $SO(3)$, i.e.,
the one with integer values of $n$  \cite{wyb}, has been used.

Again, the breakdown of the symmetry under the transformation $n
\rightarrow -(n+1)$ is built in the geometrical structure of the
$E(3)$ group: the breakdown of the $n \rightarrow -(n+1)$ symmetry
is nothing but the breakdown of the $x_{4}$ axis reversal symmetry
(breakdown of the loop-antiloop symmetry), i.e. of time-reversal
symmetry when $x_{4}$ is considered to be the time variable, which
brings us back to the analysis of subsection 3.1 (see also Section
2). Also in the present case, the $SO(4)$ contraction to $E(3)$
manifests itself in local observations and the $x_{4}$ axis
orientation then gets "locked". Which amounts to the loss of
symmetry of the solutions under the $n \rightarrow - (n+1)$
exchange.

\section{Loop algebras as extension of Euclidean algebras}\label{s6}

Infinite-dimensional algebras and in particular the so-called
``loop-algebras'' are particularly interesting and useful
algebraic structures. It is well known that they can be
constructed on some finite-dimensional group as SU(1,1)
\cite{zachos} and many applications exist which exploit such a
feature.

In view of the strict relation between the Euclidean groups, the
Bessel functions and the dissipation/amplification processes we
have discussed in the previous Sections, it is also much
interesting to consider some topological properties of Bessel
functions in connection with loop-algebras.

It is known  \cite{mekhfi} that the Bessel functions describe
solutions with different Pontryagin number in the punctured plane
$R^{2}/(0)$, where the elements of the homotopy group, $\Pi_n$,
may be represented by differential operators acting on analytic
functions:
\be
 \Pi_n \,\equiv \, \frac{\partial^n}{\partial z^n} ~, \quad n \in N ~,
\lab{A1}\ee
with
 $\Pi_n \cdot \Pi_m\ =  \  \Pi_{n+m}$ and $n$ is the loop number around
 the hole. It is possible to distinguish two
 different kinds of behavior, corresponding to two different functions:
\be
 \frac{\partial^n}{z\partial z^n} \ \varphi_{m}(z)\ = \ (-)^m \
 \varphi_{m+n}(z)~,
 \quad \varphi_{m}(z) = \frac{J_m(z)}{z^m} \label{A2}\ee
and
\be
 \frac{\partial^n}{z\partial z^n} \ \psi_{m}(z)\ = \
 \psi_{m-n}(z)~,
\quad \psi_{m}(z) = {z^m}{J_m(z)} ~, \label{A3}\ee
so that on $\varphi$, $\Pi_n $ acts in counter-clockwise way while
on $\psi$ it acts in  clockwise way. $J_m(z)$ is the planar Bessel
function (Bessel function of integer order).

Eqs.(\ref{A2}) and (\ref{A3}) are the well known differential
formulae for the planar Bessel functions  \cite{Abramowitz};
analogous formulae are true for the functions
\be \varphi_{m}(z) = {j_m(z)}{z^{-m}}~, \qquad \psi_{m}(z) =
{z^{(m+1)}}{j_m(z)} ~, \label{A4}\ee
where the $j_m$ are the spherical Bessel functions  \cite{mekhfi}.

Notice that while for the planar Bessel functions, the ``raising"
and ``lowering" functions coincide for $m=0$ (no loops), this is
not true with the spherical Bessel functions.

These topological properties, in the planar and in the spherical
case, of the $\varphi$ and $\psi$ functions readily remind us of
the $x_3$- and $x_4$-reversal symmetry breakdown discussed in
Section 3.

Thus, in view of the topological properties of the Bessel
functions and, on the other hand, of their relation with the
Euclidean groups, it is remarkable that the algebra $e(3)$ can be
related to a particular structure of loop-algebras. This goes as
follows.

Let us focus
our attention in particular on the Virasoro algebra which
plays a central role in the conformal field theories.
We will show that in analogy with the rotation algebras, it is
possible to follow a contraction procedure which maps the Virasoro
algebra into a sort of generalization of the Euclidean algebra
$e(3)$. This result derives from a general construction based on
the so called graded contraction method \cite{demontigny, moody}.

The commutation relations of the Virasoro algebra $\cal L$ of
central charge $c$ ($c$ commuting with all the $T$'s) are
\be [T_{n}, T_{m}]\,=\, (n-m) T_{n+m} \,+\, \frac{c}{12}(n^{3}-n)
\delta_{n+m,0} ~, ~~~m,n ~\in  {\cal {\bf Z}} ~. \label{6.1}\ee
The ${\cal {\bf Z}}_2$-grading of the algebra consists in dividing
the set of the $T_{n}$ generators into an even set ${L_{0}} \equiv
\{ A_{n},c \}$ and an odd set $L_{1} \equiv \{B_{n}\}$, with
\be A_{n} = \frac{1}{2}\left(T_{2n} + \frac{c}{8}\delta_{n,0}
\right)~, ~~~ B_{n} = \frac{1}{2}T_{2n+1}~, \label{6.2} \ee
so that $\cal L$ = $L_{0}\ \bigoplus \ L_{1}$ and
\be [L_{0},L_{0}] \subseteq L_{0}~, ~~~~ [ L_{0},L_{1}] \subseteq
L_{1}~, ~~~~[ L_{1},L_{1}] \subseteq  L_{0}~. \ee

The commutation relations of the graded generators are explicitly
given by \cite{gromov}
\be [A_{n}, A_{m}]\,=\, (n-m) A_{n+m} \,+\, \frac{2c}{12}(n^{3}-n)
\delta_{n+m,0} ~,\label{6.2a}\ee
 \be
[B_{n}, B_{m}]\,=\,  (n-m) A_{n+m+1} + \frac{2c}{12}(n - \half)(n
+ \half)(n + \frac{3}{2})\delta_{n+m+1,0}~, \label{6.3}\ee
\be [A_{n}, B_{m}]\,=\, (n-m-\half) B_{n+m}~. \label{6.4}\ee
Eq. (\ref{6.2a}) shows that $\{ A_{n},c \}$ is again a Virasoro
algebra but with central charge $2c$.

We can then consider the ${{\cal {\bf Z}}_2}$-graded contraction
of the algebra (\ref{6.2a}) -(\ref{6.4}) \cite{gromov}:

\be [A_{n}, A_{m}]\,=\, (n-m) A_{n+m} \,+\, \frac{2c}{12}(n^{3}-n)
\delta_{n+m,0} ~, \label{6.2}\ee
 \be
[B_{n}, B_{m}]\,=\, 0~, \label{6.3}\ee
\be [A_{n}, B_{m}]\,=\, (n-m-\half) B_{n+m}~. \label{6.4}\ee

Our remark is now that, in the centerless case $(c = 0)$,  the
$A_{0}$ and $A_{\pm 1}$ generators close the algebra isomorphic to
$so(3) \sim su(2)$ and that the set of these three generators and
the operators $B_{-\half}$, $B_{\half}$ and $B_{-\frac{3}{2}}$
close the $e(3)$ isomorphic algebra. This is shown by setting:
\bea & M_{+} \equiv \ A_{1}~, \quad M_{-} \equiv \ A_{-1}~, \quad
M_{3} \equiv \ i A_{0}~,  \nonumber \\
&  P_{+} \equiv \ B_{\half}, \quad P_{-} \equiv \
B_{-\frac{3}{2}}~, \quad P_{3} \equiv \ i B_{-\half} ~,
\label{6.5}\eea
where the $M$s and $P$s generators satisfy the commutation relations
(\ref{1.9}).

This result has a general extension, i.e. the  algebra ${\cal
E}_{n} \equiv  \{A_{0}, A_{\pm n}\}\ \bigoplus \ \{B_{-\half},
B_{\pm n-\half} \}\ $ reproduces the $e(3)$ algebra for each
integer value of $n$, provided the following positions are
assumed:
\bea & M_{+} \equiv \ \frac{1}{n}A_{n}~, \quad M_{-} \equiv
\frac{1}{n} A_{-n}~, \quad
M_{3} \equiv \ \frac{i}{n} A_{0} ~, \nonumber \\
&  P_{+} \equiv \ B_{n-\half}~, \quad P_{-} \equiv \
B_{-n-\frac{1}{2}}~, \quad P_{3} \equiv \ i B_{-\half} ~.
\label{6.6}\eea

As final remark we notice that the $e(2)$-algebra can be obtained
as a subalgebra of (\ref{6.6}) by choosing $A_{\pm n} = 0$, for
non-zero values of $n$.

Our conclusion is that the extension of the Virasoro algebra by
means of its ${{\cal {\bf Z}}_2}$-grading  with the subsequent
step of the ${{\cal {\bf Z}}_2}$-graded contraction appears as a
$n$-graded hierarchy of Euclidean algebras. In view of the results
obtained in the previous Sections, we see that an interesting
relation emerges between the couple of damped/amplified parametric
oscillators graded by $n$ and the loop algebras considered in the
present Section.

Exhibiting the link between these mathematical structures appears
in itself interesting. However, work is still needed in order to
fully recognize its physical meaning. Further study in such a
direction is in our plans.

\section{Final remarks and conclusions}\label{s4}

In this paper we have discussed the relation between the
(spherical as well as the planar) Bessel equation and the
dissipation/amplification processes. Specifically, we have shown
that, after suitable re-parametrization, the breakdown of the $n
\rightarrow -(n+1)$ ($n \rightarrow - n$)
symmetry of the original spherical (planar)
Bessel equation corresponds to the breakdown of
time-reversal symmetry (the emergence of the {\it arrow of time})
in the manifold of the solution (the Bessel functions) $\{ J_{i}
\}$, $i = n, -(n+1)$ ($i = n, - n$). We have also shown that
the mathematical
structure through which such a breakdown is realized is the
mechanism of group contraction leading to the Euclidean groups
$E(3)$ and $E(2)$, whose representations, as well known, can be
constructed by means of the spherical  and the  planar Bessel
functions, respectively. For the spherical case,
the $n \rightarrow -(n+1)$ symmetry of
the Bessel equation reflects the degeneracy, for each $n$, between
the two independent representations of $SO(3)$, $D^{n}$ and
$D^{-n-1}$. The role of the contraction mechanism is the one of
removing such a degeneracy. Then, the equivalence between the two
representations $D^{n}$ and $D^{-n-1}$ is broken and different
parametric oscillators correspond to each of them. In the sector
$\{ i=n \}$ ($\{i = -(n+1) \}$,) solutions with $l = -(n+1)$ and
with $l = n$ correspond to damped or amplified (amplified or
damped) solutions, respectively. In the large $n$ limit
($\omega_{n} \rightarrow \omega_{0}$) these solutions are each the
time-reversed of the other one and in that limit the two sectors
$\{ J_{i} \}$, $i = n, -(n+1)$ are mapped one into the other. We have
seen that, {\it mutata mutandis}, similar considerations also hold for
the planar case.  The
bottom of the representations ($n = 0$, or $n=-1$ and in the
spinor representation $n = \pm \half$) corresponds to the
undamped/un-amplified harmonic oscillator.

Summarizing, we can rephrase our finding in the following way: the
effect of the group contraction (removal of the $D^{n}$ -- $D^{-n-1}$
degeneracy) is to ``split" the original Bessel equation,
conveniently reparametrized, into a couple of damped/amplified
oscillators for each $n$.

We also observe that the fact that the symmetry of the Bessel
equation is not the symmetry of the solutions reminds us of the
phenomenon of the spontaneous breakdown of  symmetry in Quantum
Field Theory (QFT) which occurs when the {\it continuous} symmetry
of the dynamical equations is not the symmetry of the physical
vacuum. In the Bessel equation case the {\it discrete}
time-reversal symmetry is broken. However, since time-reversal
symmetry breakdown manifests itself in the appearing of
dissipation/amplification phenomena, also continuous time
translational symmetry is broken (energy non-conservation for the
damped (or amplified) system), making the resemblance with
the QFT case stronger. One further point of contact with the
quantum case is in the fact that in QFT the effects of spontaneous
breakdown of symmetry are solely observable in the contraction
limit, namely at the observation scale (``locality") which is
always "small" with respect to the system volume (the infinite
volume limit) \cite{CP76}.  Also in the present case time-reversal
symmetry is broken at local, observational level and there
dissipation/amplification phenomena become observable phenomena.

It is interesting to observe that, as well known  \cite{talman},
the Weyl-Heisenberg  algebra  of the quantization procedure is
also obtained through the group contraction of the $SO(3)$ group
(or, in other instances  \cite{gerry}, of the $SU(1,1)$ group).

In this paper our analysis has been limited to the classical
sector. The problem of quantization of the damped harmonic
oscillator, a prototype of dissipative  system, has been
extensively studied in a number of papers \cite{Dekker,FT},
\cite{QD} - \cite{BPV}, \cite{AV2}. We recall  that in order to
perform the canonical quantization of the simple damped harmonic
oscillator, one "doubles" the system by introducing the
time-reversed copy of it (the companion amplified oscillator)
\cite{FT,QD}. Canonical quantization is in fact only possible for
Hamiltonian systems, and the doubling of the dissipative
(non-hamiltonian, open) system amounts to  "closing" it by
inclusion of the environment (represented by the doubled,
amplified oscillator, indeed). At a classical level, however, it
is possible to solve the damped oscillator equation and, in
principle, to treat the dissipative system by ignoring the
environment (i.e. there is no necessity of introducing the
amplified oscillator as instead required in the canonical
quantization procedure). Remarkably, from our discussion in the
present paper it emerges that Bessel equation may represent both
the damped and the amplified oscillator as an inseparable
"double-face" unity also at the classical level.

Finally, we have also considered some topological properties of
the Bessel functions in connection with Virasoro-like
loop-algebras and we have presented some preliminary results on
the relation between such algebras and the algebras $e(2)$ and
$e(3)$. More work is still necessary in this direction in view of
the physical relevance of infinite dimensional algebras.

\ack

 We are glad to acknowledge partial
financial support by the ESF Program COSLAB, INFN and INFM.

$$   $$


\appendix
\section{Bessel equation vs damped parametric oscillators}



\smallskip

We show in the following that, under several choices of suitable
transformations, the Bessel-like equation
\be J^{''}  \ +   \frac{\alpha}{\eta} \ J^{'}  +   \left(1 -
\frac{\beta^{2}}{\eta^{2}} \right)\ J = \ 0
\label{a.1}\ee
describes damped parametric oscillators with constant damping. In
Eq.(\ref{a.1}) $\alpha=1$ or $2$, $\beta$ is an arbitrary real
constant and $^\prime$ indicates the derivative with respect to
the independent variable $\eta$.

By using the position:
\be J(\eta)  \ = x(t) \ f(\eta), \qquad t=t( \eta), \label{a.2}\ee
Eq.(\ref{a.1}) turns into:
\be \fl f(t^{\prime})^{2}\left\{\stackrel{..} x + \stackrel{.}x
\left[2 \frac{f^{\prime}}{f}\frac{1}{t^{\prime}} + \frac{
t^{\prime \prime}}{ (t^{\prime})^{2}} + \frac{\alpha}{\eta
t^{\prime}}\right] + x \left[\frac{f^{\prime \prime}}{f} + \frac{
\alpha}{\eta} \frac{f^{\prime}}{f} +1 -\frac{\beta^{2}}{ \eta^2}
\right]\frac{1}{( t^{\prime})^2} \right\} = 0. \label{a.3}\ee
Eq.(\ref{a.3}) describes a damped parametric oscillator, with
constant damping $\gamma$ and frequency $\omega$, provided the
following two conditions are satisfied:
\be 2\ \frac{f^{\prime}}{f} \frac{1} { t^{\prime}} \ + \ \frac{
t^{\prime \prime}}{ (t^{\prime})^{2}} \ + \ \frac{ \alpha}{ \eta
t^{\prime}} \ = \ \gamma, \qquad \qquad \gamma \ \ {\rm constant},
\label{a.31}\ee
i.e.
\be f = f_{0}\ e^{(\gamma/2)t} \ \eta^{-\alpha/2}\
(t^{\prime})^{-\half},
\label{a.4}\ee
and
\be \left[ \frac{f^{\prime \prime}}{f}  \ + \ \frac{ \alpha}{
\eta} \frac{f^{\prime}}{f} +1 -\frac{ \beta^{2}}{ \eta^2}
\right]\frac{ 1}{ ( t^{\prime})^2}\ = \ \omega^{2}.
\label{a.5}\ee
Let us consider conditions (\ref{a.31})--(\ref{a.5}) for specific
choices for $t(\eta)$:

\smallskip
A) ~~~$t= \eta^{l}$, ~~ $l$ integer
\smallskip

\be   f \propto \exp(\frac{\gamma \eta^{l}}{2}) \
\eta^{\frac{-\alpha}{2} + \frac{(1-l)}{2}}~.
\label{a.6}\ee

In the simple case $l =1$ ($t=\eta$), $\omega$ is given by


\be
\omega^{2}= (\frac{\alpha}{2}- \frac{\alpha^2}{4}-
\beta^{2})\frac{1}{t^{2}} + (\frac{\gamma^{2}}{4}+1)~.
\label{a.6a}\ee
Note that for $\frac{\alpha}{2}-\frac{\alpha^2}{4}= \beta^{2}$,~
$\omega^2$ gets constant values.
\smallskip

\smallskip
B) ~~~$t= \exp{(q\eta)}$, ~~ $q$ integer
\smallskip

\be  f \propto \eta^{-\alpha/2}\
\exp\left[-\frac{q\eta}{2}+\frac{\gamma}{2}e^{(q\eta)} \right]~,
\label{a.7}\ee
with
 \be \omega^{2}=\left(\frac{\gamma}{2}\right)^{2} +
\left(\frac{1}{q t \eta}\right)^{2}
[\frac{\alpha}{2}(1-\frac{\alpha} {2}) - \beta^{2}]
+\frac{1}{t^{2}}\left[\frac{1}{4}+ \frac{1}{q^{2}}\right]~,
\label{a.8}\ee
that is, for $\alpha=1$ and $\beta=\half$:
\be \omega^{2}= \left(\frac{\gamma}{2}\right)^{2}+
\left[\frac{1}{4}+ \frac{1}{q^{2}}\right]{\frac{1}{t^{2}}}.
\label{a.9}\ee

\smallskip
C) ~~~$t= \ln{(q\eta)}$, ~~ $q$ integer
\smallskip

\be  f \propto \eta^{(\gamma+1-\alpha)/2}~,
\label{a.10}\ee
and
\be \omega^{2}\propto {\cal A} + {\cal B} e^{2t}, ~~~~{\cal
A},{\cal B} ~ constant.
\label{a.11}\ee

\smallskip
D) ~~~~$t= \sin{(q\eta)}$, ~~ $q$ integer
\smallskip

\be f \propto \exp{\frac{\gamma \sin{q\eta} }{2}} \eta^{-\alpha/2}
(\cos{q\eta})^{-\half} ~,
\lab{6.12}\ee
and, for $\alpha=1, \beta= \half$,
\be \omega^{2}\propto {\cal A} + \frac{{\cal
B}}{(\cos{q\eta})^{2}}+ \frac{{\cal C}}{(\cos{q\eta})^{4}}~.
\lab{6.13}\ee

It is straightforward that (almost) all the forms for $\omega$ are
possible. Vice-versa, under a suitable transformation, it  is
always possible to turn the equation for a parametric oscillator,
with constant damping $\gamma$ and frequency $\omega$, into a
Bessel-like equation of type (\ref{a.1}).



\end{document}